\documentclass[]{article}
\usepackage{multicol}
\usepackage{apj}
\def\lesssim{\mathrel{\hbox{\rlap{\hbox{\lower4pt\hbox{$\sim$}}}\hbox{$<$}}}}
\def\gtrsim{\mathrel{\hbox{\rlap{\hbox{\lower4pt\hbox{$\sim$}}}\hbox{$>$}}}}
\newcommand{\mincir}{\raise -2.truept\hbox{\rlap{\hbox{$\sim$}}\raise5.truept
\hbox{$<$}\ }}
\newcommand{\magcir}{\raise -2.truept\hbox{\rlap{\hbox{$\sim$}}\raise5.truept
\hbox{$>$}\ }}
\newcommand{\siml}{\raise -2.truept\hbox{\rlap{\hbox{$\sim$}}\raise5.truept
\hbox{$<$}\ }}
\newcommand{\simg}{\raise -2.truept\hbox{\rlap{\hbox{$\sim$}}\raise5.truept
\hbox{$>$}\ }}

\newcommand{\be}{\begin{equation}} 
\newcommand{\ee}{\end{equation}}
\newcommand{\ba}{\begin{eqnarray}} 
\newcommand{\ea}{\end{eqnarray}}
\newcommand {\h} {$h^{-1} \, Mpc \ $} 
\newcommand {\hh} {$h^{-1} \, Mpc$} 
\newcommand {\ks} {$km~s^{-1}  \ $} 
\newcommand {\kss} {$km~s^{-1}$} 
\newcommand {\msun} {$h^{-1} \   M_{\odot} \ $} 
\newcommand {\msunn} {$h^{-1} \   M_{\odot}$}

\begin{document}


\vspace{15mm}
\begin{center}
\uppercase{
The Observational Mass Function \\ of Loose Galaxy Groups}\\
\vspace*{1.5ex}
{\sc Marisa Girardi$^{1}$ and Giuliano Giuricin$^{1,2}$}\\
\vspace*{1.ex}
{\small
$^1$
Dipartimento di Astronomia, Universit\`{a}
degli Studi di Trieste, Via Tiepolo 11, I-34131 Trieste, Italy\\
$^2$SISSA, via Beirut 4, I-34014 Trieste, Italy\\
E-mail: girardi@ts.astro.it, giuricin@sissa.it\\}
\end{center}

\begin{abstract}
We analyze the three catalogs of nearby loose groups by Garcia
(1993). She identified groups in a magnitude--limited redshift galaxy
catalog, which covers about $\sim 2/3$ of sky within $cz =5500$ \kss,
by using two methods, a percolation and a hierarchical method. The
free parameters of the group-selection algorithms were tuned to obtain
similar catalogs of groups. The author also proposed a third catalog
of groups defined as a combination of the two.  Each catalog contains
almost 500 groups.

In agreement with previous works on earlier catalogs, we find that
groups can be described as collapsing systems. Their sampled size is
in general considerably larger than their expected virialized region.
We compute the virial masses and correct them by taking into account
the young dynamical status of these groups.  We estimate corrected
group masses, $M$, for two reference cosmological models, a flat one
with a matter density parameter 
$\Omega_0=1$ and an open one
with $\Omega_0=0.2$.  For each of the three catalogs we calculate the
mass function.

We find that the amplitude of the mass function is not very sensitive
to the choice of the group-identification algorithm. The number
density of groups with $M> 9 \times 10^{12}$ \msunn, which is the
adopted limit of sample completeness, ranges in the interval
$1.3$--$1.9 \times 10^{-3} h^{3}Mpc^{-3}$ for
$\Omega_0=1$, and it is about a factor of $15\%$ lower for
$\Omega_0=0.2$. The mass functions of the
hierarchical and combined catalogs have essentially the same shape,
while the mass function of the percolation catalog shows a flattening
towards large masses.  However, the difference decreases if we do not
consider the most massive groups, for which reliable results come from
galaxy cluster studies.

After having estimated the mass contained within the central,
presumably virialized, regions of groups by adopting a reduction in
mass of $\sim 30$--$40\%$, we do a comparison with the results coming
from the virial analysis of nearby rich clusters (Girardi et
al. 1998a).  All three group mass functions turn out to be a smooth
extrapolation of the cluster mass function at $M<4 \times
10^{14}$\msunn, which is the completeness limit of the cluster sample.
The resulting optical virial mass function of galaxy systems, which
extends over two orders of magnitude, is fitted to a Schechter
expression with a slope of $\sim -1.5$ and a characteristic mass
of $M^*\sim 3\times 10^{14}$\msunn.  We also verify that our group
mass function reasonably agrees with the Press--Schechter predictions
of models which at large masses describe the virial mass
function of clusters.

%

\vspace*{6pt}

\noindent
{\em Subject headings: }
galaxies: clusters: general - cosmology: observations -
cosmology: theory - large scale structure of universe.

\end{abstract}

\begin{multicols}{2}
\section{INTRODUCTION}
Most galaxies in the local universe belong to loose galaxy groups.
Groups seem to be the natural continuation of galaxy clusters at
smaller mass scales. Indeed, there is a continuity of properties from
rich clusters to poor clusters and to groups (e.g., Ramella, Geller,
\& Huchra 1989; Burns et al. 1996; Mulchaey \& Zabludoff 1998;
Ramella et al. 1999; Girardi, Boschin, \& da Costa 2000).

Zabludoff \& Mulchaey (1998) used multi--fiber spectroscopy to obtain
velocities for a large number of group members (i.e. 280 galaxies for
a total of 12 groups) and Mahdavi et al. (1999) measured several
hundreds of redshifts to obtain a sample of 20 groups, each one having
, on average, 30 galaxies .  For these well--sampled groups Zabludoff
\& Mulchaey (1998) and Mahdavi et al. (1999) performed refined
analyses, i.e. the rejection of interlopers, the study of the internal
galaxy distribution and velocity dispersion profiles, and the separation
of different galaxy populations (see, e.g.,  Biviano et al. 1997;
Carlberg et al. 1996, 1997b; den Hartog \& Katgert 1996; Dressler et
al. 1999; Girardi et al. 1996, 1998b; Mohr et al. 1996; Koranyi \&
Geller 2000; for recent relevant results on rich and poor
clusters). 

However, all these analyses are so far restricted to a limited number
of groups since they require a strong observational effort.
Therefore, to analyze group dynamical properties in a statistical
sense, one must resort to wide group catalogs where groups are
extracted from three--dimensional galaxy catalogs and typically
contain $\lesssim$ five member galaxies (e.g., Huchra \& Geller 1982;
Tully 1987; Ramella et al. 1999).

Here, we focus our attention on the determination of group mass
function from wide catalogs of nearby loose groups.  The observational
determination of group mass function is plagued by several
problems. Some of them concern the estimate of mass and are mainly due
to the small number of group members and to uncertainties in the
dynamical stage.  In fact, although group cores are virialized or
close to virialization (Zabludoff \& Mulchaey 1998), the size of
groups identified in three--dimensional galaxy catalogs, i.e. $\sim
0.5$--$1$ \hh, is appreciably greater than their expected virialized
region, i.e. $\sim 0.2$--0.4 \h for systems with a line--of--sight
velocity dispersion of 100--200 \ks (according to the relations found
for galaxy clusters, e.g. Carlberg, Yee, \& Ellingson 1997; Girardi et
al. 1998b).  Indeed, there is a strong indication that these groups
are not virialized systems over the whole sampled region, but can be
rather described as being in a phase of collapse (e.g., Giuricin et
al. 1988; Mamon 1994; Diaferio et al. 1993).  Therefore, usual
estimates of velocity dispersion and virial mass are not easily
connected to physical quantities such as group potential and mass.

The small number of data and the uncertainties on dynamical status
prevent one to use refined methods to reject interlopers in each
individual group (e.g., Zabludoff \& Mulchaey 1998, Mahdavi et
al. 1999).  Instead, one must rely on member galaxies as assigned by
the group-selection algorithm, while checking a posteriori the
presence of spurious groups in a statistical sense (e.g.,  Ramella,
Pisani, \& Geller 1997; Diaferio et al. 1999). Indeed the results
could depend on the choice of the group-selection algorithm and its
free parameter (e.g., Pisani et al. 1992 -- hereafter P92; Ramella et
al. 1997).  For instance, Frederic' s (1995b) analysis of cosmological
N--body simulations suggested that the estimated median mass depends
on the algorithm and that the resulting bias is sensitive to the depth
of the galaxy survey. However, even the analysis of simulated groups
is not an easy task and, indeed, the results on mass can depend on the
treatment of halos (cf. Frederic 1995b).

A further uncertainty is connected to cosmic variance.  In fact, group
catalogs are recovered from local galaxy catalogs which may not be
fair samples of the universe. 

In view of these difficulties, few statistical distributions of group
dynamical properties are available in the literature and they are
often discrepant.  The cumulative distributions of internal
velocity--dispersion, as computed by Moore, Frenk, \& White (1993) and
by Zabludoff et al.  (1993), are strongly discrepant (the number
densities of groups with line--of--sight velocity dispersion larger
than 200 \ks differ by a factor of 100, see Fig.~6 of Fadda et
al. 1996 for a comparison). Moreover, analyzing nearby groups ($cz\le
2000$ \kss) of three different group catalogs, P92 found a significant
dependence of the distribution of mass and other dynamical group
parameters on the group-identification algorithm.

The availability of new group catalogs has prompted us to derive a new
group mass function, whose connection with the recent determination of
the optical virial mass function of nearby rich galaxy clusters
(Girardi et al. 1998a, hereafter G98) deserves to be investigated.

The work by Garcia (1993, hereafter G93), who constructed two group
catalogs using two different identification algorithms (the
percolation and hierarchical ones) and proposed a third catalog which
is a combination of the two, represents a good data base for facing
the effect of identification algorithms.  So far, G93 catalogs are the
largest catalogs of groups presently published. They are largely
superior to those analyzed by P92 both for the number of groups
(450--500 groups for each of the three catalogs) and the encompassed
volumes ($\sim 2/3$ of sky, $cz\le 5500$ \kss).  Moreover, these group
catalogs were selected from the same parent galaxy sample, thus
allowing us to better investigate on the effects due to differences in
the selection algorithm.  Furthermore, the improved statistics in the
high-mass range (less than ten groups analyzed by P92 have masses
larger than $10^{14}$\msunn) permits an interesting comparison with
cluster mass function and a determination of the virial mass function
over an unprecedently large range of masses.

In \S~2 we briefly describe the data. In \S~3 we calculate group
masses. In \S~4 and 5 we compute group mass function and verify its
stability, respectively. In \S~6 we compare the results of groups and
clusters, recovering the mass function of galaxy systems for a mass
range which extends over two orders of magnitude. In \S~7 we give our
discussions.  In \S~8 we summarize our results and draw our
conclusions.

Throughout the paper, errors are given at the $68\%$ confidence level and
the Hubble constant is $H_0=100\ \rm{h}\ $Mpc$^{-1}$ \kss.
\section{DATA SAMPLE}
We analyze the loose group catalogs constructed by G93. These groups
were identified by using galaxies (within $cz=5500$ \kss) belonging to
the subsample of the Lyon--Meudon Extragalactic Database which is
nearly complete down to the limiting apparent magnitude $B_0=14$, the
total blue magnitude corrected for Galactic absorption, internal
absorption and K--dimming.  G93 used two methods in group
construction: a percolation method (hereafter $P$, derived from the
friends--of--friends method presented by Huchra \& Geller 1982) and a
hierarchical method (hereafter $H$, derived from that of Tully 1980,
1987).  Each method gives one catalog. The $P$ and $H$ catalogs
contain 453 and 498 groups of at least three members, respectively.
In particular, G93 tuned the free parameters of the methods so as to
obtain the best compromise between the stability of group membership
and the similarity of the two group catalogs (Garcia, Morenas, \&
Paturel et al. 1992).

Then G93 combined together the two catalogs in order to obtain the
final catalog (hereafter $G$ catalog, 485 groups) defined 
as the catalog which contains only the groups which were found in part
in both catalogs. For these final groups only the galaxies in common
were kept as group members. If some groups of  a catalog
(in most cases the $P$ catalog) turned out  to be divided into two 
or more groups in the other catalog (in most cases the $H$ catalog),
the smallest systems were kept for the final catalog.

In our work, as already suggested by the author, we do not consider
galaxies added afterwards (flagged by ``+'') which have no known
magnitude or which are not fully satisfying the selection criteria. We
refer to the original paper for more details on group catalogs.

We exclude from our analysis groups with $cz\le500$ \ks because, when
the velocity becomes low, its random component dominates and the
velocity is no longer a reliable indication of the distance.

The samples we analyze consist of 446, 490, and 476 groups ($P$, $H$,
and $G$ groups, respectively).
\section{COMPUTING GROUP MASSES}
The virial mass of a relaxed galaxy system is computed as
$M_{vir}=\sigma^2 R_V/G$ where $R_V$ is the virial radius of the
system and $\sigma$ is the velocity dispersion of member galaxies
(Limber \& Mathews 1960).  Other usual mass estimators, e.g. the
projected and the median mass mass estimators (Bahcall \& Tremaine
1981; Heisler, Tremaine, \& Bahcall 1985), give similar results as
shown by P92 (cf. also Perea, del Olmo, \& Moles 1990). A more serious
problem comes from the fact that one must assume that, within each
group, mass distribution follows galaxy distribution (e.g., Merritt
1987). This assumption is shown to be enough reliable for galaxy
clusters both from optical, X--ray, and gravitational lensing analyses
(see Girardi et al. 1998b, Lewis et al. 1999 and references therein)
and, for likeness, the same assumption can be made for galaxy groups.
 
We compute the above virial parameters from the observed projected
positions in the sky and the line--of--sight velocities of the member
galaxies. In fact, for a spherical system, the parameters $R_V$ and
$\sigma$ are linked to their observational counterparts as
$\sigma=\sqrt{3} \sigma_{v}$ and $R_V=(\pi/2) R_{PV}=(\pi/2)
N_m(N_m-1)/\Sigma_{i>j}R_{ij}^{-1}$, where $\sigma_v$ is the
line--of--sight velocity dispersion, $R_{PV}$ is the projected virial
radius, $N_m$ is the number of group members and $R_{ij}$ the galaxy
projected distances.  In particular, we estimate the "robust" velocity
dispersion by using the biweight estimator for rich groups (member
number $N_m\ge15$) and the gapper estimator for poorer groups (ROSTAT
routines by Beers, Flynn, \& Gebhardt 1990). Beers et al.  (1990) have
shown the superiority of their techniques in terms of efficiency and
stability when treating systems with a small number of members
(cf. also Girardi et al. 1993). In our case (see also Mahdavi et
al. 1999) we verify that the distributions of robust and traditional
estimates of velocity dispersion are not different according to the
Kolmogorov--Smirnov test (hereafter KS-test, e.g.  Ledermann 1982).
We apply the relativistic correction and the usual correction for
velocity errors (Danese, De Zotti, \& di Tullio 1980).  In particular,
for each galaxy, we assume a typical velocity error of $30$ \ks based
on the average of errors estimated in RC3 catalog from optical and
radio spectroscopy (de Vaucouleurs et al.  1991).

Note that our estimates of the virial parameters do not require any
luminosity--weighting procedure. Indeed, it was shown that the values
of the virial masses are largely insensitive to different weighting
procedures (e.g., Giuricin, Mardirossian, \& Mezzetti 1982; P92) in
agreement with evidences of velocity equipartition (e.g., Giuricin et
al. 1982).

In order to take into account the dynamical state of groups we use the
method proposed by Giuricin et al. (1988). This method is based on the
classical model of the spherical collapse where the initial density
fluctuation grows, lagging behind the cosmic expansion when it breaks
away from the Hubble flow, and begins to collapse; then a relaxation
process sets in (e.g., Gott \& Rees 1975).  The time evolution curves
$A(t)$, which is the ratio between the absolute values of the kinetic
and potential energies, given by the dynamical model, are the starting
point for the determination of the evolutionary stage. The evolution
curve used by Giuricin et al. (1988) has been derived from numerical
simulations of systems composed of 15 point masses with a
Schechter--type mass function (Giuricin et al. 1984; the limits of
this model are discussed in \S 7.1)

According to the above method, the value of $A$, which is needed to
recover corrected masses as $M=(1/2A)(\sigma^2 R_V/G)$, can be
inferred from the estimate of the presently observed virial crossing
time $t_{cr}=(3/5)^{3/2}R_V/\sigma$.  One can also derive the value of
$\tau$, which is the elapsed time since fluctuations started growing
(here in units of the crossing time at the virialization,
$t_{cr}^{v}$).  In particular, the precise values of $A$ and $\tau$
depend on the background cosmology.  Here, we estimate corrected group
masses for two reference cosmological models, a flat one with
$\Omega_0=1$ for the matter density parameter and an open one with
$\Omega_0=0.2$ (cf. also P92).

For each catalog, in Table~1 we give the number of groups (Col.~2) and
median group properties: the mean redshift (Col.~3); the
line--of--sight velocity dispersion, $\sigma_v$ (Col.~4); the projected
virial radius, $R_{PV}$ (Col.~5); the crossing time, $t_{cr}$
(Col.~6); the virial mass, $M_{vir}$ (Col.~7); the value of $\tau$ and
the corrected virial mass, $M$, for the $\Omega_0=0.2$ model (Cols.~8
and 9) and for the $\Omega_0=1$ model (Cols.~10 and 11).

\end{multicols}
\begin{minipage}{4cm}
\renewcommand{\arraystretch}{1.2}
\renewcommand{\tabcolsep}{1.2mm}
TABLE 1\\
{\sc Group Properties\\}
\footnotesize
%
%

\begin{tabular}{lcccccrcrcc}
\hline \hline
\multicolumn{1}{c}{Cat.}
&\multicolumn{1}{c}{$N$}
&\multicolumn{1}{c}{$z$}
&\multicolumn{1}{c}{$\sigma_v$}
&\multicolumn{1}{c}{$R_{PV}$}
&\multicolumn{1}{c}{$H_0t_{cr}$}
&\multicolumn{1}{c}{$M_{vir}$}
&\multicolumn{1}{c}{$\tau_{0.2}$}
&\multicolumn{1}{c}{$M_{0.2}$}
&\multicolumn{1}{c}{$\tau_1$}
&\multicolumn{1}{c}{$M_1$}
\\
\multicolumn{1}{c}{}
&\multicolumn{1}{c}{}
&\multicolumn{1}{c}{}
&\multicolumn{1}{c}{$(km\,s^{-1})$}
&\multicolumn{1}{c}{$(h^{-1}Mpc)$}
&\multicolumn{1}{c}{}
&\multicolumn{1}{c}{$(h^{-1}M_{\odot})$}
&\multicolumn{1}{c}{$(t_{cr}^{v})$}
&\multicolumn{1}{c}{$(h^{-1}M_{\odot})$}
&\multicolumn{1}{c}{$(t_{cr}^{v})$}
&\multicolumn{1}{c}{$(h^{-1}M_{\odot})$}
\\
\multicolumn{1}{c}{(1)}
&\multicolumn{1}{c}{(2)}
&\multicolumn{1}{c}{(3)}
&\multicolumn{1}{c}{(4)}
&\multicolumn{1}{c}{(5)}
&\multicolumn{1}{c}{(6)}
&\multicolumn{1}{c}{(7)}
&\multicolumn{1}{c}{(8)}
&\multicolumn{1}{c}{(9)}
&\multicolumn{1}{c}{(10)}
&\multicolumn{1}{c}{(11)}
\\
\hline
$G$        &476&0.0097&118&0.59&0.23& $8.4E12$&5.43&$10.7E12$&5.22&$12.7E12$\\
$H$        &490&0.0096&108&0.62&0.26& $6.6E12$&5.32& 9.8$E12$&5.11&$11.8E12$\\
$P$        &446&0.0095&151&0.62&0.20&$14.8E12$&5.51&18.1$E12$&5.31&$20.3E12$\\
\hline
\end{tabular}

\end{minipage}
\begin{multicols}{2}

From the values of $\tau$ ($<2\pi$) we infer that most groups are in
the phase of collapse and not yet virialized in agreement with
previous analyses of earlier catalogs (e.g., Giuricin et al.  1988;
P92).  Moreover, $P$ groups result to be more evolved than $H$ groups.
The resulting estimate of the corrected mass is larger than the virial
mass by a factor of $20$--$60\%$, depending on the group catalog and
on the assumed background cosmology. Since, for each of the three
catalogs, $t_{cr}$ does not correlate with $M_{vir}$, one could apply
the same percent correction to $M_{vir}$ of all groups to obtain
corrected masses (as inferred from Table~1, e.g. multiplying by
$12.7/8.4=1.41$ in the case of $G$ groups and $\Omega_0=1$). The mass
distribution resulting from the application of an average correction
is indistinguishable (according to the KS-test) from the application
of individual corrections for each group.
  
The distribution of corrected group masses is only slightly dependent
on the cosmological environment; in fact, the two mass distributions
computed for $\Omega_0=0.2$ and $\Omega_0=1$ do not differ
significantly (according to the KS-test). Therefore, hereafter, if not
explicitly said, we consider the $\Omega_0=1$ case only.
\section{GROUP MASS FUNCTION}
In order to compute a reliable group mass function, MF, we avoid
strongly obscured regions by considering only groups with galactic
latitude $|b|\ge 20^{\circ}$ (i.e. within a solid angle of $\omega=8.27$
sr). We consider 409, 381, and 344 $G$, $H$, and $P$ groups,
respectively.

To obtain the true spatial number density, the selection effects of
the galaxy catalog, which is magnitude limited, must be taken into
account.  Following Moore et al.  (1993), we weight each group by
using the magnitude of its third brightest galaxy which allows the
inclusion of the group itself in the catalog. We weight each group by
$w=1/\Gamma_{max}$ where $\Gamma_{max}$ is the maximum volume, within
the catalog velocity cut--off, out to which the group can be seen:
\ba
\Gamma_{max}=\left(\frac{\omega}{3}\right)\cdot\left(v_{lim}H_0^{-1}\right)^3 \left[1-\frac{3}{2}\left(1+q_0\right)\frac{v_{lim}} {c}\right]
\nonumber\\
-\left(\frac{\omega}{3}\right)\cdot\left(500 \cdot H_0^{-1}\right)^3 \left[1-\frac{3}{2}\left(1+q_0\right)\frac{500}{c}\right],
\ea
\noindent where $v_{lim}$ is the smaller one between 5500 \ks and the
maximum recession velocity at which the third brightest galaxy in the
group would be brighter than the magnitude limit; 
$c$ is the speed of light and $q_0$ the
deceleration parameter.

By plotting the weights vs. group masses (cf. Figure~1, left panels),
we note that groups generally lie in a well--defined region of the
$w-M$ plane with the exception of a few groups which fall very far
away from the other points.  Therefore, in order to avoid problems of
instability in the resulting MF, for a few groups we recompute the
weights according to the following procedure. In each mass range (of a
unity in logarithmic scale) we compute the mean and s.d. of $log\,w$
and substitute the values which are three s.d. far away from the mean
with the corresponding mean values.  We change the value of weights
for three, four, and three groups in $G$, $H$, and $P$ catalogs,
respectively (cf. Figure~1, right panels).

\includegraphics{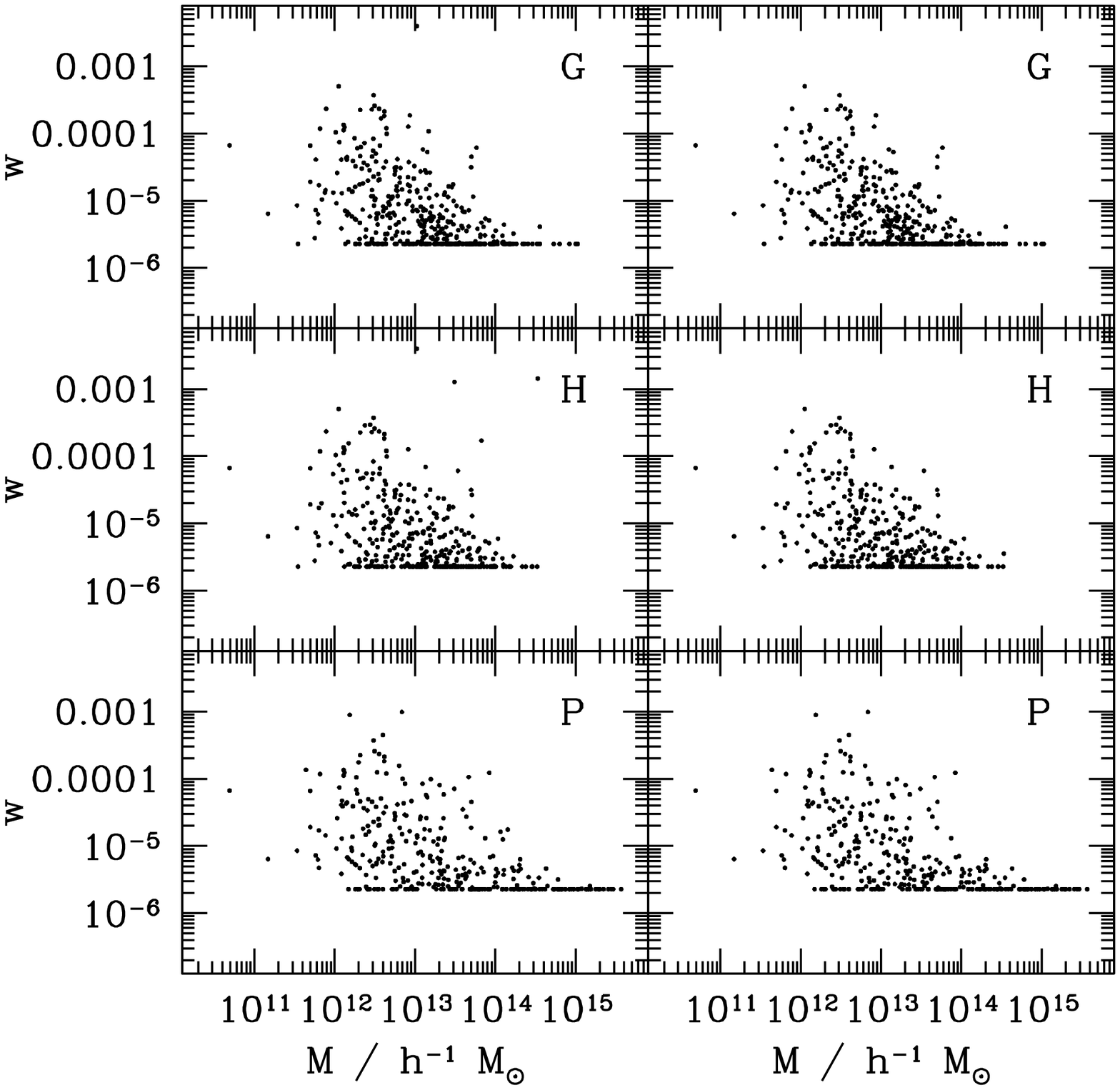}
$\ \ \ \ \ \ $\\
\vspace{8.2truecm}
$\ \ \ $\\
{\small\parindent=3.5mm {Fig.}~1.---
For the three catalogs, left panels show that
groups generally lie in a well--defined region of the weight--mass
plane.  Right panels show the same, after recomputing the value of
weights for a few groups which lie far from this region (see text).
}
\vspace{5mm}

Then, since the parent galaxy sample was found to have a redshift
incompleteness of $\sim 10\%$ (Marinoni et al. 1999), the resulting
group densities are enhanced by the same factor. This  is a rough 
correction. As discussed in \S~5, a more refined correction 
which takes into account the number of group members 
would give similar results. 

The comparison between different catalogs is shown in Figure~2.  In
the low--mass range the three MFs show large fluctuations (well over
the estimated Poissonian errors) and a tendency towards flattening,
which suggests problems of incompleteness.  Our results for $M \gtrsim
9\times 10^{12}$ \msunn, which we assume as our completeness limit,
come from 230, 207, and 214 groups for $G$, $H$, and $P$ catalogs,
respectively.  The determination of global group number density is
quite stable, since the density of groups with $M> 9 \times
10^{12}$\msun is $1.4$, $1.3$, and $1.9 \times 10^{-3}
(h^{-1}Mpc)^{-3}$ for $G$, $H$, and $P$ catalogs, respectively.

However, Figure~2 shows that $P$ catalog gives a flatter MF with
respect to $G$ and $H$ catalogs.  In fact, according to the KS--test,
the MF of $P$ groups strongly differs from that of $G$ and $H$ groups
(at a c.l. greater than 98\%), while there is no 
difference between $G$ and $H$ groups.  However, the difference
becomes smaller if we exclude large masses and, for instance, we do
not find any significant difference if we exclude $ M > 4 \times
10^{14}$\msunn, a range where reliable results come from cluster
analysis.

In the case of the $\Omega_0=0.2$ model we find similar results,
except for the fact that, since estimated masses are smaller, the
density of groups with $M> 9 \times 10^{12}$\msun is somewhat lower
than in the $\Omega_0=1$ case, i.e. $1.2$, $1.0$, and $1.7
\times 10^{-3} (h^{-1}Mpc)^{-3}$ for $G$, $H$, and $P$ catalogs,
respectively.

\includegraphics{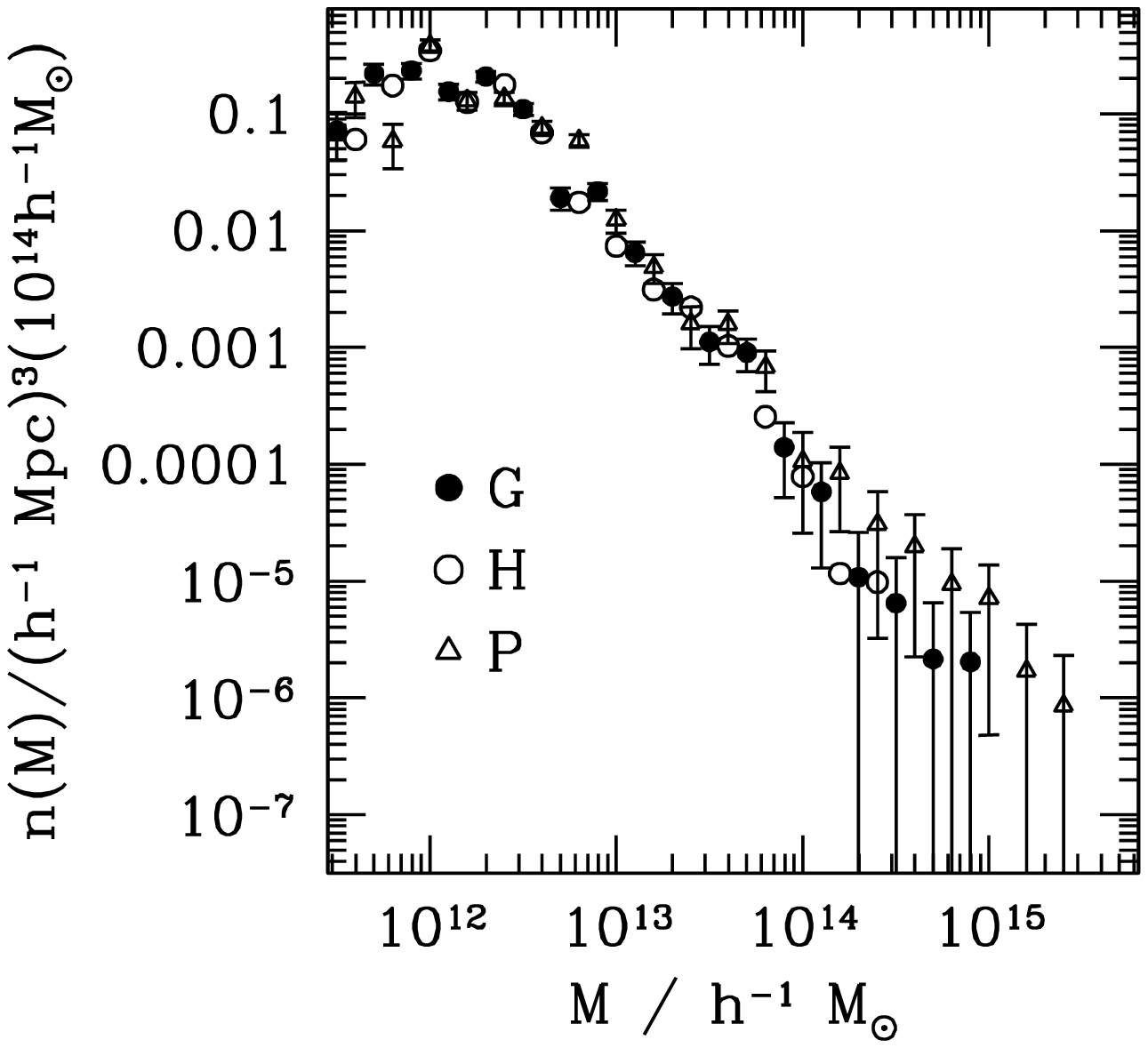}
$\ \ \ \ \ \ $\\
\vspace{7.truecm}
$\ \ \ $\\
{\small\parindent=3.5mm {Fig.}~2.---
We compare mass functions as computed from the
three catalogs. $G$, $H$, and $P$ groups are denoted by closed
circles, open circles, and triangles, respectively.  To avoid
confusion we give error bars ($1\sigma$ Poissonian uncertainties) only
for $G$ and $P$ groups.
}
\vspace{5mm}

\section{ON THE STABILITY \\ OF  GROUP MASS FUNCTION}

\includegraphics{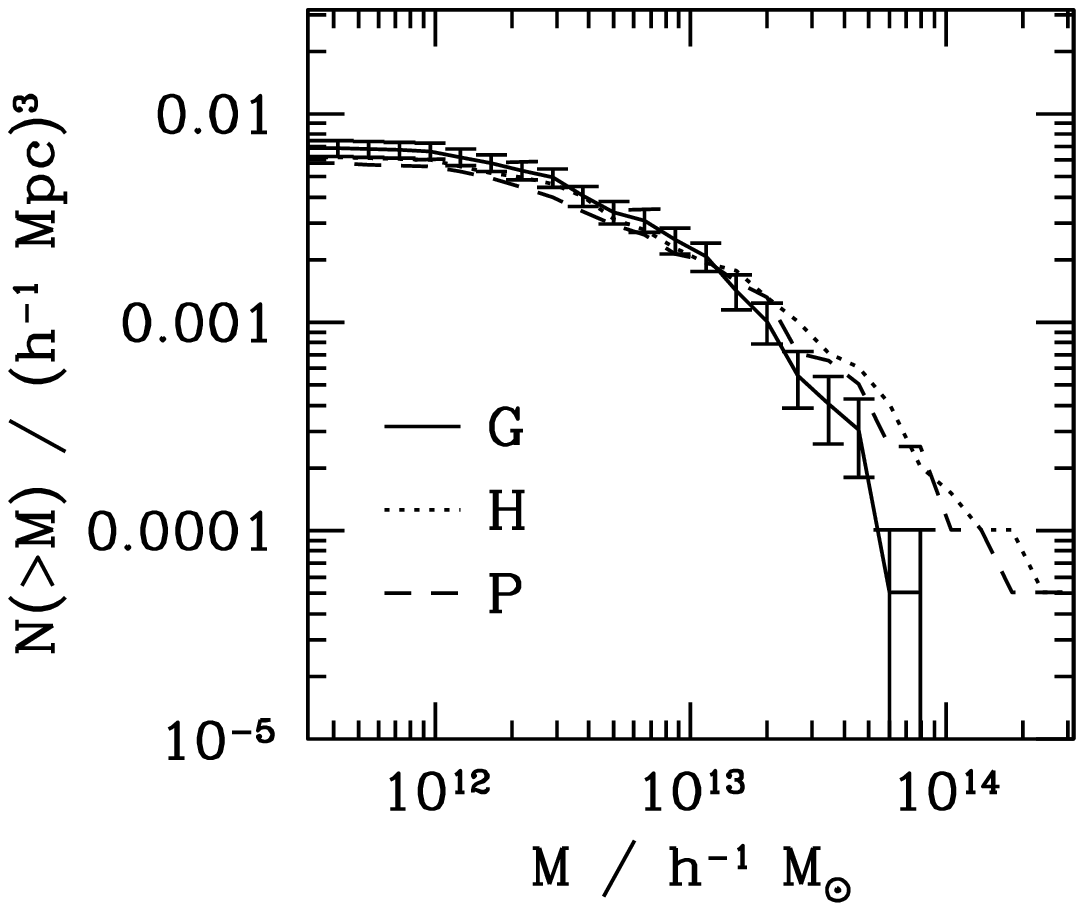}
$\ \ \ \ \ \ $\\
\vspace{5.5truecm}
$\ \ \ $\\
{\small\parindent=3.5mm {Fig.}~3.---
The cumulative mass function for the nearby
groups of $G$, $H$, and $P$ catalogs (with $cz\le 2000$ \ks).  Error
bars represent $1\sigma$ Poissonian uncertainties.
}
\vspace{5mm}

First we consider groups with $cz\le2000$ \kss. In particular, we have
136, 124, and 117 $G$, $H$, and $P$ nearby groups, respectively,
with galactic latitude $|b|\ge 20^{\circ}$. For each catalog, the
observed group population in the nearby subsamples can be considered a
good representation of the total population because we do not see any
significant trend between mass and distance. Therefore, the
nearby groups can be considered a complete sample, except for the
less massive ones, which could not be identified out to the distance
limit of 20 \h  since their third brightest galaxy is not bright
enough. If we assume that the nearby groups form a complete sample, we
can directly compute the MF for each of the three catalogs. There is
no significant difference among the three mass distributions
(according to the Kruskas--Wallis test, e.g., Ledermann 1982, cf.
Figure~3).  Moreover, apart from the density of low-mass groups,
$M\sim 10^{12}$ \msunn, where the completeness of the nearby groups is
questionable, there is a good agreement between the number densities
estimated from the whole catalogs with those from the nearby ones
(cf. Figure~4).  This result reassures us of the reliability of our
weighting procedure (cf. \S~4).

\includegraphics{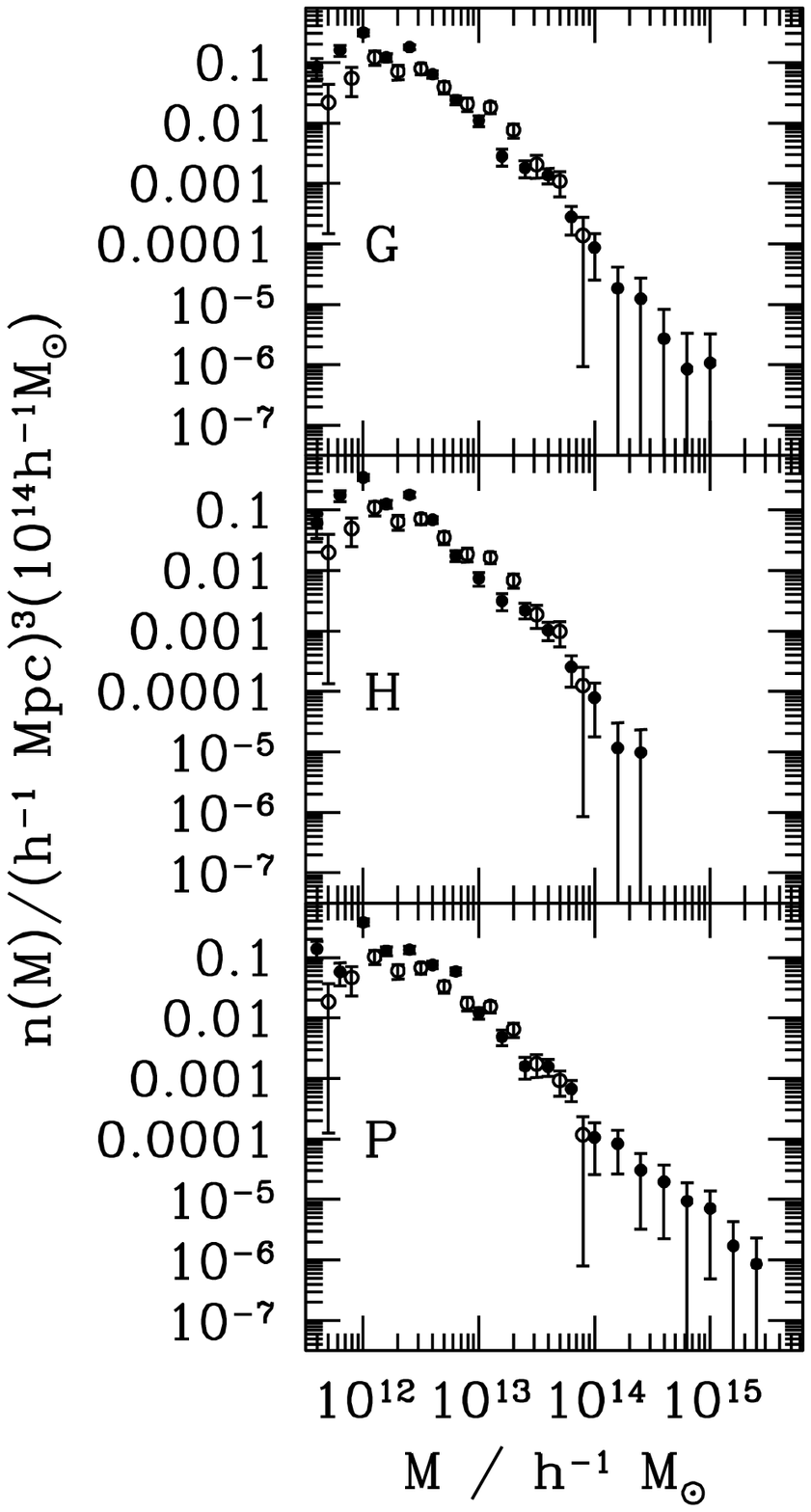}
$\ \ \ \ \ \ $\\
\vspace{12.7truecm}
$\ \ \ $\\
{\small\parindent=3.5mm {Fig.}~4.---
For the three catalogs we compare mass
functions as computed from all the groups and the nearby groups
(closed and open circles, respectively).  Error bars represent
$1\sigma$ Poissonian uncertainties.
}
\vspace{5mm}

The physical reality of the detected groups is often discussed in the
literature. In particular, the efficiency of the percolation algorithm
has been repeatedly checked through cosmological N--body simulations
(e.g., Nolthenius \& White 1987; Moore et al. 1993; Nolthenius,
Klypin, \& Primack 1994; Frederic 1995a,b; Nolthenius, Klypin, \&
Primack 1997; Diaferio et al. 1999) and geometrical Monte--Carlo
simulations (Ramella et al. 1997). These computations show that an
appreciable fraction of the poorer groups, those with $N_m<5$ members,
is false (i.e. unbound density fluctuations), whereas the richer
groups almost always correspond to real systems (e.g., Ramella et
al. 1995; Mahdavi et al. 1997).  By following the results of Ramella
et al. (1997; cf. also Diaferio et al. 1999) we reduce the weights of
groups with $N_m=3$ and $4\le N_m \le 5$ members by $70\%$ and $20\%$,
respectively. Figure~5 shows the effect on the cumulative MF of $P$
groups. The MF at large masses is quite stable and, however, the
number density at the completeness limit of $ 9\times 10^{12}$ \msun
could  be overestimated only by a factor of $\lesssim
50\%$.

\includegraphics{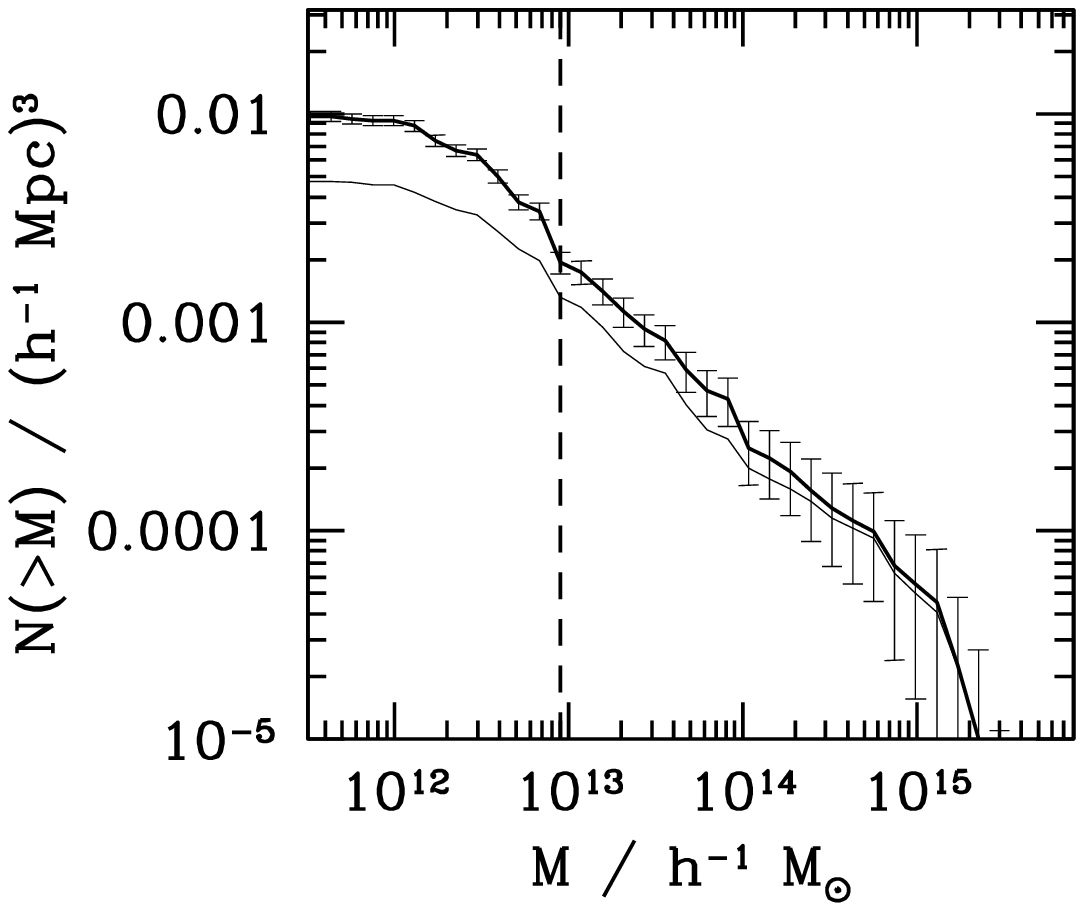}
$\ \ \ \ \ \ $\\
\vspace{5.3truecm}
$\ \ \ $\\
{\small\parindent=3.5mm {Fig.}~5.---
We show the effect of the presence of poor
groups on the cumulative mass function of $P$ groups.  The thin line
indicates the mass function obtained when the weight of poor groups is
suitably reduced (see text).  Error bars represent $1\sigma$
Poissonian uncertainties.  The mass completeness limit is indicated by
the vertical dashed line ($M>9 \times 10^{12}$ \msunn).
}
\vspace{5mm}

Galaxy density is known to show significant fluctuations around the
mean density. Redshift surveys reveal voids of sizes up to 50 \h and
large bidimensional sheets (e.g., de Lapparent, Geller, \& Huchra
1986; Geller \& Huchra et al. 1989; Vettolani et al. 1997).  Indeed,
there is also an indication of a local underdensity of the above size
(e.g., Zucca et al. 1997; Marinoni et al. 1999).  Since group catalogs
are obtained from galaxy catalogs, one could suspect that the
amplitude of the group MF we estimate in the local universe is far
from being a fair value. In order to shed light on this point, we use
the groups identified in the northern Center for Astrophysics redshift
survey (CfA2N) by Ramella et al. (1997). The CfA2N covers a smaller
solid angle (1.2 sr) than the survey analyzed by G93, but it is deeper
($cz\le 12000$ \kss) and contains the Great Wall, a very overdense
structure . From Ramella et al. (1997) we take group dynamical
quantities, $M_{vir}$ and $t_{cr}$, and apply the same procedure
outlined in \S~3 and 4 for the evolutionary correction and for the
computation of weights (we take the luminosity of the third brightest
galaxies from Ramella et al. 2000, in preparation). The resulting MF
is shown in Figure~6.  The number density of the CfA2N groups with $M>
9 \times 10^{12}$\msun lies within the range of values we found for G93
catalogs. In conclusion, although the CfA2N groups come from a quite
different volume, their MF is similar to the MF of the G93 groups.
This result agrees with that by Ramella et al. (1999) who found that
different group catalogs, which sample different volumes, give similar
velocity--dispersion distributions.

\includegraphics{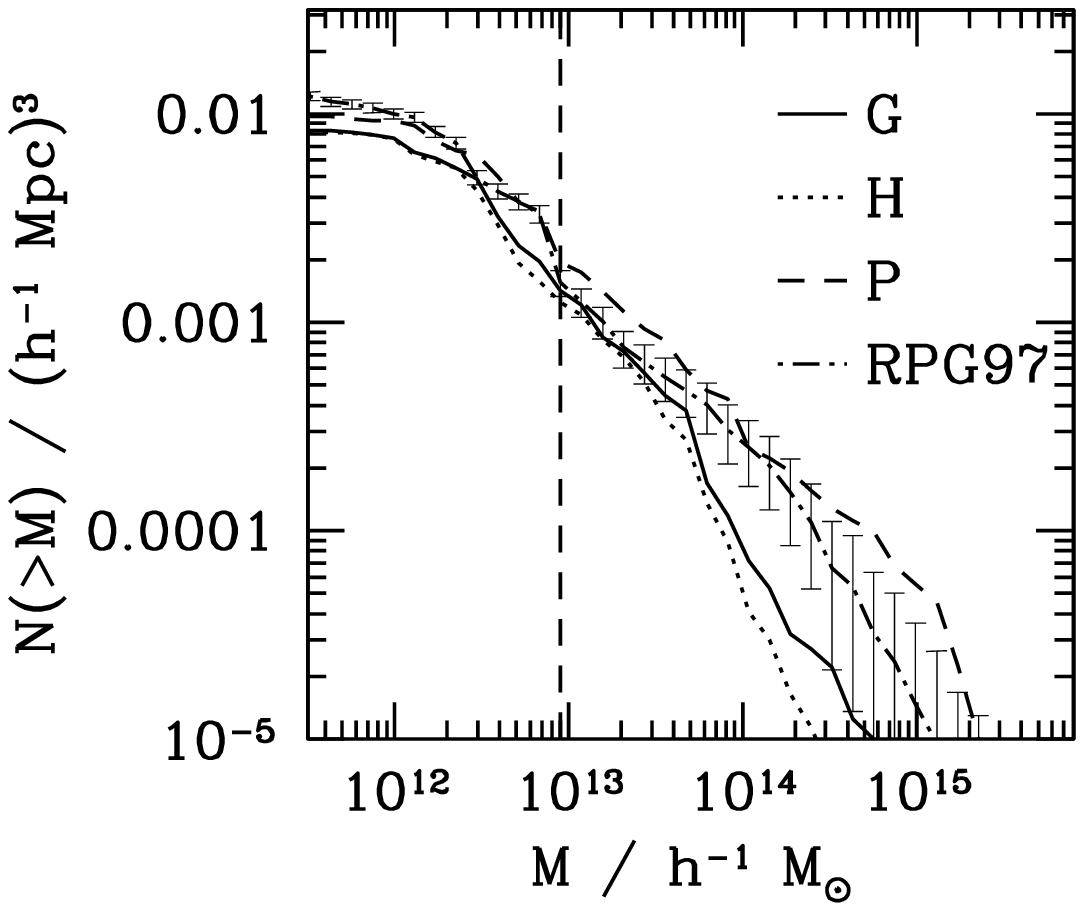}
$\ \ \ \ \ \ $\\
\vspace{5.3truecm}
$\ \ \ $\\
{\small\parindent=3.5mm {Fig.}~6.---
We compare the cumulative mass function of the
CfA2N groups by Ramella et al. (1997; RPG97) with that of $G$, $H$,
and $P$ groups.  Error bars represent $1\sigma$ Poissonian
uncertainties.  The mass completeness limit is indicated by the
vertical dashed line ($M>9 \times 10^{12}$ \msunn).
}
\vspace{5mm}

In \S~4 in order to take into account the 10\% redshift incompleteness
of the parent galaxy sample (Marinoni et al. 1999), we have applied a
rough, very small upward correction by 10\% to the group densities. A
more refined correction should take into account the number of members
in each group.  For instance, let's adopt the extreme view that the
incompleteness of parent galaxy sample does not affect the density of
groups with more than three members, but affects only groups with
three members. In the worst case, out of 100 groups (i.e. 300
galaxies, of which 30 are missed) one misses 30 groups. According
to this kind of incompleteness we recompute the MFs for the three
catalogs. The differences in densities for groups with $M> 9 \times
10^{12}$ are very negligible for $G$ groups and at most amount to
$\lesssim10\%$, which is within the errors, for $H$ and $P$ groups.
\section{GROUP VS. CLUSTER MASS FUNCTION}
Contrary to the case of groups, recent determinations of the
distribution of internal velocity dispersions for nearby rich clusters
well agree within the errors (cf.  Figures~6 and 9 of Fadda et
al. 1996).  The cluster MF based on optical virial masses has recently
been presented by G98. In particular, masses were computed within a
radius enclosing the region where clusters are virialized. The
resulting MF is reliably estimated for masses larger than $4\times
10^{14}$\msun with $\sim 50$ nearby clusters available for this mass
range.

We must take into account that the groups examined here extend outside
their virialized regions for a meaningful comparison of group masses
with cluster masses as well as with Press \& Schechter (1974,
hereafter PS) predictions, which hold for virialized objects
(e.g., Eke, Cole, \& Frenk 1996; Lacey \& Cole 1996; Borgani et
al. 1999).

By using eq.~3 of Ramella et al. (1997), we estimate that $P$ groups
has a limiting number density contrast of $\sim 70$, while for $H$
groups, detected imposing a luminosity density threshold (Gourgoulhon,
Chamaraux, \& Fouqu\`e 1992; G93), we compute a luminosity density
contrast of $\sim 40$.  These calculations use the Schechter
luminosity function (1976) with a slope of $-1.1$, a normalization
factor of $0.014$ $h^3\,Mpc^{-3}$, and a characteristic magnitude of
$M_B^*=-20.0-5\cdot \,log\,h$, as well as the galaxy blue luminosity
density, $\sim 4.5\times 10^8$ $L_{\odot}\,h^3\,Mpc^{-3}$, as obtained
by Marinoni et al. (1999) for the same parent galaxy sample.  The
estimated values of density contrast should be considered as rough
estimates also in view of the difficulties in obtaining group catalogs
with a constant density contrast (e.g., Nolthenius \& White 1987).

In order to compute the matter density contrast, $\delta \rho/\rho$,
from the above galaxy density contrast, $(\delta \rho/\rho)_g$, one
should know the biasing factor $b= (\delta \rho/ \rho)_g/(\delta\rho/
\rho)$.  Herebelow we compute the value of bias as $b=1/\sigma_8$
where $\sigma_8$, the r.m.s. mass density fluctuation in spheres of 8
\h radius, is estimated for different values of $\Omega_0$ according
to the G98 relation found by comparing PS predictions to the cluster
MF (cf. their eq.~4).  Accordingly, we adopt $b=1/0.60\sim 1.7$ and
$b=1/1.23\sim 0.8$ in the case of the $\Omega_0=1$ and the
$\Omega_0=0.2$ models, respectively.

The resulting $\delta \rho/ \rho$ for $P$ and $H$ groups (41 and 24,
respectively, for $\Omega_0=1$; 88 and 50 for 
$\Omega_0=0.2$) are much smaller than the values of $\sim 180$
and $\sim 550$ expected within the virialized region for
$\Omega_0=1$ and $\Omega_0=0.2$, respectively (e.g., Eke et al. 1996).

After assuming that groups have a common radial profile (Fasano et
al. 1993), we can roughly estimate the fraction of the mass contained
in the virialized region.  For each of the three catalogs, Figure~7
plots the cumulative distributions of the galaxy distances from the
group (biweight) center combining together data of all groups. To
combine the galaxies of all groups we divide each galaxy distance to
the projected virial radius, $R_{PV}$, of its group. Moreover, in
order to better show the behavior of galaxy density, we also normalize
the distances to the mean $<R_{max}/R_{PV}>$ of the catalog, where the
maximum radius, $R_{max}$, is the projected distance from the group
center of the most distant galaxy, and we normalize the numbers of
objects, $N$, to that contained within $<R_{max}/R_{PV}>$.
  
From Figure~7 one can infer the fraction of the number of galaxies
(i.e. the fraction of group mass if galaxies trace mass) contained
within each radius (i.e. how density changes with radius).  We are
interested in determining the radius (and the corresponding galaxy
number/group mass) for which one obtains a density enhancement which
is large enough to reach the density contrast expected at
virialization. In particular, in the case of $\Omega_0=1$ a reduction
in radius of $\sim 50\%$ (i.e. in number/mass of $\sim 30\%$) is
enough to reach the virialization density.  In fact, from Figure~7 one
infers that $70\%$ of galaxies are contained within about half of the
radius ($0.545$ for $P$ groups; $0.49$ for $H$ groups). Therefore,
considering only these central group regions, the density increase is
$=0.7/(0.545^3)\sim4.3$ for $P$ groups and $0.7/(0.49^3)\sim5.9$ for $H$
groups; in other words, the resulting density contrast in a
$\Omega_0=1$ universe is $\sim 41\times4.3\sim 180$ for $P$ groups and
$\sim 24\times5.9\sim 140$ for $G$ groups (these values are comparable
to the virialization density contrast). In view of the inherent
approximations, we simply adopt a reduction of group masses by $30\%$
for all groups in each of the three catalogs.  In the case of
$\Omega_0=0.2$, similar arguments lead us to adopt a reduction in
radius of $\sim 60\%$ (i.e. in number of $\sim 40\%$) to reach the
virialization density.

In the following comparison with clusters we will reduce group masses
by $30\%$ and $40\%$ for the $\Omega_0=1$ and the $\Omega_0=0.2$
cases, respectively.  We remark that this kind of correction is taken
to be equal for all groups since we assume a common typical spatial
distribution of member galaxies (Fasano et al. 1993) and we do not
consider each individual evolutionary state (note that the crossing
time does not show any correlation with group mass, cf. the end of
\S~3).

\includegraphics{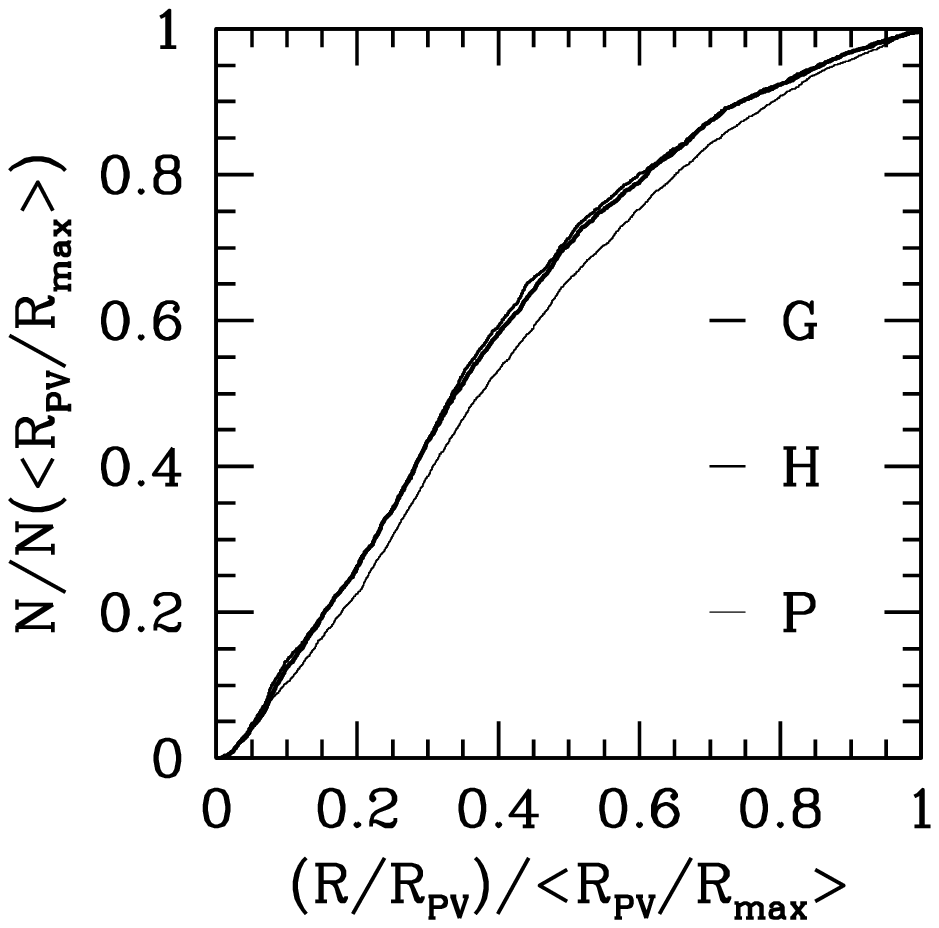}
$\ \ \ \ \ \ $\\
\vspace{5.3truecm}
$\ \ \ $\\
{\small\parindent=3.5mm {Fig.}~7.---
For each of the three catalogs we give the
cumulative distribution of the galaxy distances from group center. To
combine the galaxies of all groups we divide each galaxy distance to
the projected virial radius, $R_{PV}$, of its group.  Moreover, we
also normalize the distances to the mean $<R_{max}/R_{PV}>$ of the
catalog, and normalize the numbers of objects, $N$, to that contained
within $<R_{max}/R_{PV}>$.  The distributions of $G$ and $H$ are
nearly overlapping.
}
\vspace{5mm}

Figure~8 shows the group and the cluster MFs. The $P$ group MF gives
too high values at high masses when compared with cluster results,
while $G$ and $H$ catalogs do not contain very massive groups.
Therefore, the differences among group catalogs in describing the MF
at high masses (cf. \S~4) lead to strong uncertainties just where
cluster data are available. However, for $M<4 \times 10^{14}$\msun all
the three group MFs can be regarded as a smooth extrapolation of the
cluster MF.

To provide a phenomenological fitting to the data which takes into
account the effect of mass uncertainties we use the Schechter
expression (1976).  A theoretical, PS approach is instead discussed at
the end of this section.

Following a maximum--likelihood approach we fit the Schechter
expression  for our MF on the whole mass range (groups with
$9\times 10^{12}$\msun $<M< 4 \times 10^{14}$ \msun and clusters with
$M>4 \times 10^{14}$ \msunn):
\begin{equation}
n(M)=n^* \left(\frac{M}{M^*}\right)^{-\alpha} e^{-M/M^*},
\end{equation}
\noindent where $n(M)$ is suitably convolved with the mass errors
$\Delta M$.  We compute uncertainties on group masses in the same way
as in Girardi et al. (1998b): we propagate statistical errors in the
estimate of the velocity dispersion $\sigma_v$ and of the virial
radius $R_{PV}$, for which errors are estimated via bootstrap and
jacknife techniques, respectively.  The resulting mass uncertainties
range from $\sim 15\%$ for clusters at $M\sim 2\times 10^{15}$ \msun
to $\sim 90\%$ for groups at $M\sim 9\times 10^{12}$ \msun (we fit
$log (\Delta M/M)= 4.548-0.355\cdot log\,M$) and we assume a lognormal
error distribution.  The effect of these uncertainties is negligible
in the cluster mass range, but it becomes significant in the range of
low mass groups.

For instance, in the case of $G$ groups ($\Omega_0=1$ model) the
resulting fitted parameters are: $n^*=(2.2\pm0.3) \times 10^{-5}$
$(h^{-1} Mpc)^{-3}$ $(10^{14}h^{-1} M_{\odot})^{-1}$,
$M^*=3.1^{+1.8}_{-1.3} \times 10^{14}$ \msunn, and
$\alpha=1.55^{+0.14}_{-0.16}$.  Table~2 gives the fit results for the
three catalogs and for both cosmological model: (Col.~3) gives the
number of groups used in this analysis; (Cols.~4, 5, and 6) give the
fitted $n^*$, $M^*$, and $\alpha$ parameters, respectively.  The
values of $\alpha$ vary in the range of $1.3$--$1.6$ and the values of
$M^*$ vary in the range of $2.7$--$3.9\times 10^{14}$\msunn.

\includegraphics{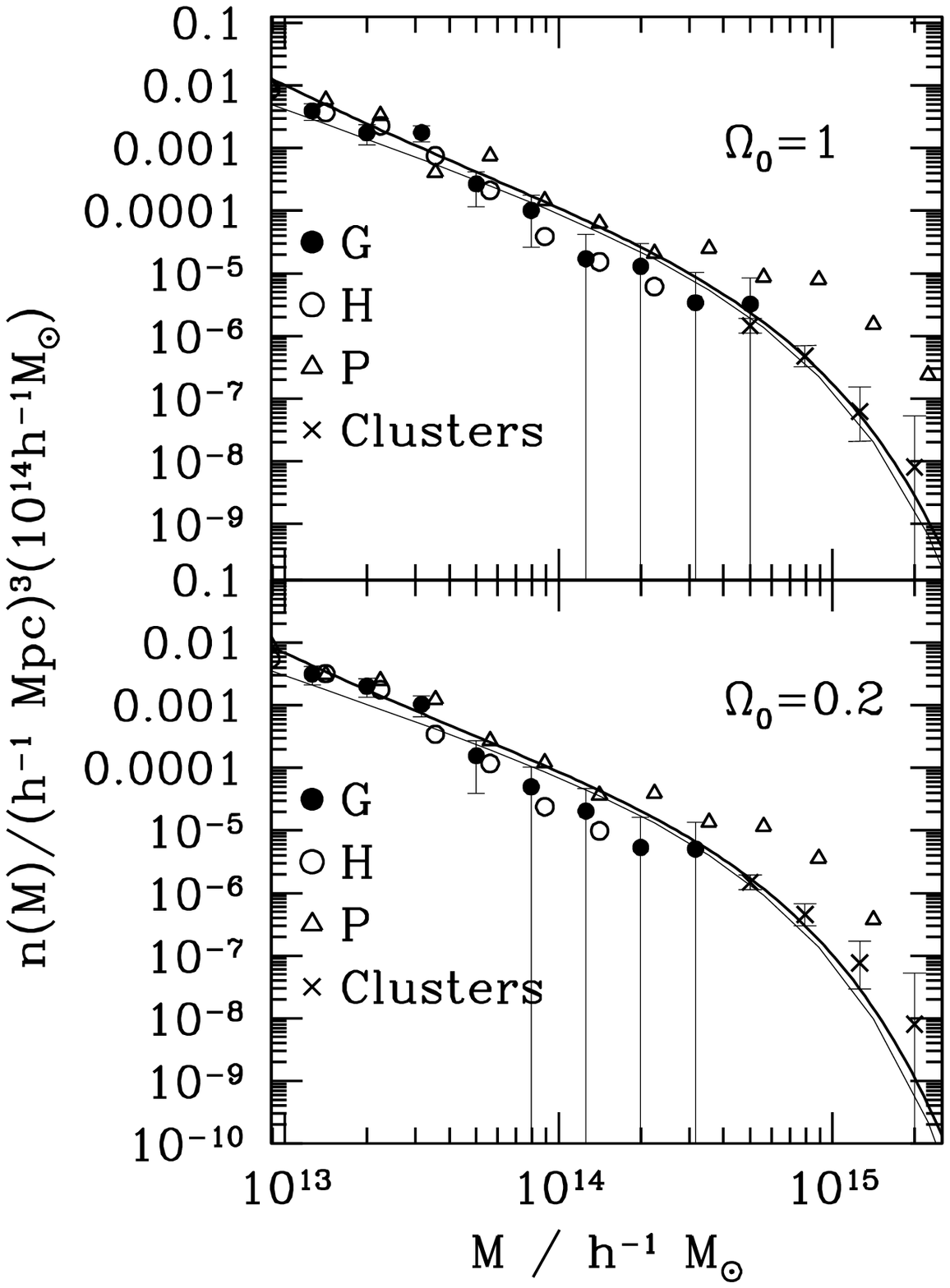}
$\ \ \ \ \ \ $\\
\vspace{11.2truecm}
$\ \ \ $\\
{\small\parindent=3.5mm {Fig.}~8.---
For the two reference cosmological models and
for each of the three catalogs we show group and mass functions, where
masses are computed within the virialized region. Cluster data come
from Girardi et al. (1998a).  We show the fitted Schechter function
and its convolution with errors (thin and thick lines, respectively)
by combining data of $G$ groups and clusters.
}
\vspace{5mm}

\end{multicols}
\begin{minipage}{9cm}
\renewcommand{\arraystretch}{1.2}
\renewcommand{\tabcolsep}{1.2mm}
TABLE 2\\
{\sc Parameters of the Mass Function Fit\\}
\footnotesize
%
%

\begin{tabular}{llcccr}
\hline \hline
\multicolumn{1}{c}{Cat.}
&\multicolumn{1}{c}{Model}
&\multicolumn{1}{c}{$N$}
&\multicolumn{1}{c}{$n^*$}
&\multicolumn{1}{c}{$M^*$}
&\multicolumn{1}{c}{$\alpha$}
\\
\multicolumn{1}{c}{}
&\multicolumn{1}{c}{}
&\multicolumn{1}{c}{}
&\multicolumn{1}{c}{$[(h^{-1} Mpc)^{-3}\,(10^{14}\,h^{-1}\,M_{\odot})^{-1}]$}
&\multicolumn{1}{c}{$(10^{14}\,h^{-1}\,M_{\odot})$}
&\multicolumn{1}{c}{}
\\
\multicolumn{1}{c}{(1)}
&\multicolumn{1}{c}{(2)}
&\multicolumn{1}{c}{(3)}
&\multicolumn{1}{c}{(4)}
&\multicolumn{1}{c}{(5)}
&\multicolumn{1}{c}{(6)}
\\
\hline 
$G$&$\Omega_0=1$  &193& 2.2$ \times 10^{-5}$&3.1&1.55\\
$H$&$\Omega_0=1$  &177& 1.0$ \times 10^{-5}$&3.9&1.64\\
$P$&$\Omega_0=1$  &153& 4.6$ \times 10^{-5}$&3.1&1.42\\
$G$&$\Omega_0=0.2$&159& 2.4$ \times 10^{-5}$&2.7&1.49\\
$H$&$\Omega_0=0.2$&147& 1.5$ \times 10^{-5}$&3.1&1.55\\
$G$&$\Omega_0=0.2$&153& 5.5$ \times 10^{-5}$&3.0&1.31\\
\hline 
\end{tabular}

\end{minipage}
\begin{multicols}{2}

The errors on the "shape" parameters $\alpha$ and $M^*$ are directly
given by the maximum likelihood method (Avni 1976).  We recover the
error on the normalization $n^*$ by considering the Poissonian error
bars associated to the global number of objects considered (and
assuming the best fit parameters for the shape).  These errors are the
formal ones and do not consider other additional effects.  The effect
of incompleteness (see the end of \S~5) is very small, being smaller
than Poissonian errors. The effect due to the presence of interlopers
is evidenced by differences in results relative to different
group-selection algorithms.  In particular, for $P$ groups, for which
the presence of spurious groups is well studied in the literature,
adopting the correction analyzed in \S~5, we find values for the slope
and characteristic mass ($\alpha=1.5$ and $M^*=3.3\times 10^{14}$\msun
for the case $\Omega_0=1$, respectively) which are within the errors,
but a $n^*$ smaller by a factor of $50\%$, e.g. $n^*=2.4 \times
10^{-5}$ $(h^{-1} Mpc)^{-3}$ $(10^{14}h^{-1} M_{\odot})^{-1}$, as
already pointed out in \S~5. However, this normalization is similar to
that coming from the $G$ catalog.

G98 applied the PS approach to constrain the cosmological parameters
from the observational cluster mass function.  The PS approach
provides fairly accurate analytical approximation to the number
density of dark matter halos of a given mass.  The halo mass which
appears in the PS formula refers to the mass contained within the
virialized region, i.e. the region with a present density contrast of
$18\pi^2=178$ (for $\Omega_0=1$, see Eke et al. 1996 for values in
different cosmologies).  Moreover, the PS mass function has been
compared with N--body simulations by several authors (e.g., Eke et
al. 1996; Lacey \& Cole 1996; Gross et al. 1998; Borgani et al. 1999;
Governato et al. 1999) and has been generally shown to provide a
rather accurate description of the abundance of virialized halos of
cluster size.

\includegraphics{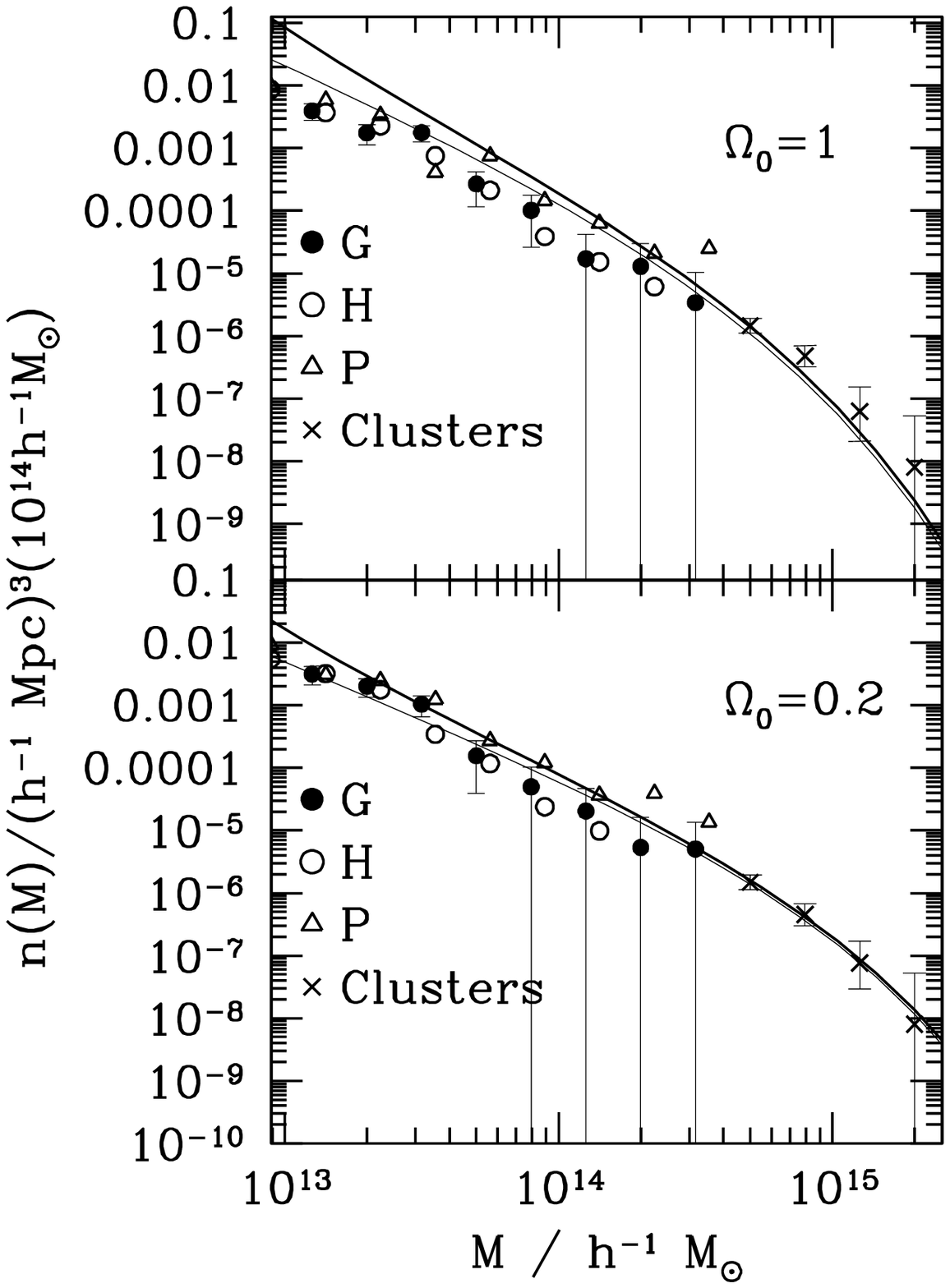}
$\ \ \ \ \ \ $\\
\vspace{11.2truecm}
$\ \ \ $\\
{\small\parindent=3.5mm {Fig.}~9.---
For the two reference cosmological models and
for each of the three catalogs we compare observational mass functions
with Press--Schechter mass functions, with and without convolving with
the uncertainties in the mass estimates (thick and thin lines,
respectively). The plotted Press--Schechter mass functions are those
found to well describe cluster data (Girardi et al. 1998a).
}
\vspace{5mm}

It is worth verifying that the present group MF, in spite of all
difficulties inherent in the analysis of groups (cf. \S 7.1), is a
good extension of the PS form fitted on clusters.  We use the same PS
approach used by G98 for describing the MF of galaxy clusters
(cf. their eq.~2 and 3 and details in \S~3). In particular, G98
recovered the relation between $\sigma_8$ and $\Omega_0$ (cf. their
eq.~4 for $\Omega_{\Lambda}=0$). Accordingly, we assume $\sigma_8=0.60$
and $=1.23$ for the $\Omega_0=1$ and the $\Omega_0=0.2$ cosmological
models, respectively, and we fix the shape parameter of the CDM--like
power spectrum (e.g., Bardeen et al. 1986) to $\Gamma=0.2$.  Figure~9
shows the comparison between the predictions of PS and the
observational mass functions.  Observational data can be described by
the PS models, except for the range of low--mass groups in the case of
$\Omega_0=1$. Although it is well documented that the PS model which
fits rich cluster data overpredicts the number density of low--mass
halos compared to the simulations (e.g., Gross et al. 1998; Governato
et al. 1999), the predicted difference is much smaller than the
difference between our observational MF and the (convolved) PS.
Unless the observed difference is due to some problems of data
incompleteness in the low-mass range, the case for the open model
seems to be preferable.  We emphasize that the comparison with PS
predictions is not done here for determining the best--fitting
cosmological model. Instead it is aimed at verifying that our
extension of the MF to group scales is reasonable.
\section{DISCUSSION}
\subsection{Group Mass Determination}
We find that G93 groups can be described as systems in a phase of
collapse, in agreement with previous results from earlier group
samples identified in redshift galaxy catalogs (e.g., Giuricin et
al. 1988; P92; Mamon 1994; Diaferio et al. 1993).  We find that the
presumably virialized region in groups is only $\lesssim$ half of the
radius sampled by galaxy data.

In order to determine the dynamical status of groups and the
corresponding mass correction we use the method of Giuricin et
al. (1988), which has been also applied by P92.  This method is based
on simple numerical simulations, in which groups are not framed within
a cosmological environment and the galaxies are represented by mass
points starting from a spherical distribution with zero velocity
(Giuricin et al. 1984).  The comparison with observations could be not
straightforward since the observed galaxies may be affected by several
environmental effects, e.g. tidal stripping, dynamical friction, and
merging events which are not taken into account by these simple
simulations.  This is connected to the validity of the crucial
hypothesis ``mass distribution follows galaxy distribution'' which is
needed also for the standard virial mass estimate (see, e.g., Merritt
1987, 1988 for clusters) and which is also used by us in considering
group masses contained within virialized regions (thus, our method of
determining the dynamical status is fully consistent with the mass
estimate itself).

Several N-body simulations of hierarchical clustering have been
performed in order to study galaxy systems in realistic situations.
However, until recently, the poor resolution of halos within dense
environments has led to soft, diffuse halos that are rapidly dissolved
by tidal forces (the so--called overmerging problem, White et
al. 1987). This makes it difficult to compare simulations with
observed group galaxies which are identified with halos.  Recent
simulations show that environmental effects can be important and
affect the structure of individual galaxies (e.g.,  Moore et al. 1996;
Tormen, Diaferio, \& Syer 1998; Colpi, Mayer, \& Governato 1998).  As
for the global distribution of galaxies within clusters, Ghigna et
al. (1998, 2000) performed very high-resolution simulations of
clusters where the overmerging seems to be globally
unimportant (the same seems to hold for small systems, see Moore et
al. 1999). Ghigna et al. found that the velocity dispersions of the
halos agree with that of the dark matter particles.  As for
spatial biasing, they found that, at an early epoch of cluster
formation, halos and dark matter have number density profiles of similar
shape, while at the final time halos are anti--biased.  In any case,
since up to now these results concern virialized systems, studies of
larger, unvirialized regions should be awaited before reaching
definitive conclusions for just forming groups.

From the observational point of view, the luminosity segregation of
galaxies in clusters, attributed to dynamical friction and merging,
was found to concern only the brightest or very bright galaxies
(Biviano et al. 1992; Stein et al. 1997).  The overall global
properties of clusters do not appear to depend on galaxy luminosities
and also for groups virial masses are largely insensitive to
luminosity weighting procedures (Giuricin et al. 1982; P92).

On the other hand, there are several evidences that different galaxy
populations (e.g., early-- and late--type galaxies, blue and red
galaxies, emission- or non-emission-line galaxies--hereafter ELGs and
NELGs) show different spatial and velocity distributions (Biviano et
al. 1997; Carlberg et al. 1996, 1997b; den Hartog \& Katgert 1996;
Adami, Biviano, \& Mazure 1998; Dressler et al. 1999; Girardi et
al. 1996, 1998b; Mohr et al. 1996; Stein 1997; Koranyi \& Geller
2000).  As shown by the analyses of velocity dispersion profiles and
spatial distribution, the galaxy component whose behavior most differs
from the norm is that of very late--type galaxies or alternatively
ELGs, being often ELGs very late types and vice-versa (Biviano et
al. 1997; Adami et al. 1998).  Biviano et al. (1997) suggested
that the dynamical state of the ELGs reflects the phase of galaxy
infall rather than the virialized condition in the relaxed cluster
core (cf. also Mahdavi et al. 1999). On the other hand, Carlberg et
al. (1997a) suggested that, although differing in their distributions,
both blue and red galaxies are in dynamical equilibrium with clusters,
cf. also Mazure et al. (2000) who explain ELG dynamics by
resorting to  more radial orbits with respect to NELGs.

As for global properties, as estimated within large samples, using
ENACS (ESO Nearby Abell Clusters Survey) data, Biviano et al. (1997)
found that the velocity dispersion and virial masses based on ELGs
are, on average, $20\%$ and $50\%$, respectively, larger than those
based on NELGs. However, due to the small fraction of ELGs ($\sim 10\%$
of cluster members) the presence of ELGs does not strongly affects the
estimate of velocity dispersion and mass of clusters.  As for groups,
we consider the 20 well-sampled groups analyzed by Mahdavi et
al. (1999). They detected no velocity dispersion segregation between
ELGs and NELGs together with a (not significant) decrease of masses by
20\%, if ELGs are excluded from the sample.  Therefore, even if global
dynamical properties based on ELGs and NELGs can be significantly
different, we expect that the possible presence of ELGs in the groups
we analyze hardly affects significantly, on average, our mass
estimates.

Finally, we note that the dynamical status of the groups analyzed as
well as the small number of galaxy members prevents us from applying
refined analyses (used for clusters and well--sampled groups), e.g.
the determination of velocity anisotropies from velocity dispersion
profiles and the Jeans equation. In fact, since the Jeans equation
rigorously holds only in regions being in dynamical equilibrium, this
analysis was generally applied to galaxy clusters (e.g., Carlberg et
al. 1996, 1997b; Girardi et al. 1998b). This kind of analysis was
also applied out to external regions of galaxy groups as a
successful  approximation (Mahdavi et al. 1999), but in any case it
requires a large number of data. In this case Mahdavi et al. combined
together data of a well--behaved subset of groups as selected by an
analysis of the velocity dispersion profiles of individual
groups.  In this sense, the approach we use should be viewed as an
alternative method for deriving masses in the case in which available
data do not allow sophisticated analyses.
\subsection{Comparison with Previous MF}
At present, P92 is the only study of the mass distribution function of
loose groups.  P92 determined the group mass function analyzing
nearby groups ($cz\le 2000$ \kss) of two catalogs based on the
percolation method (38 and 21 groups by Geller \& Huchra 1983 and
Maia, da Costa, \& Latham 1989, respectively) and one catalog based on
the hierarchical method (107 groups by Tully 1987). The three 
catalogs have parent galaxy samples which considerably differ both for
the selection criteria and the sky region covered.

P92 noted significant differences between the group MFs resulting from
the three catalogs of groups and claimed that these differences were
not due to the inhomogeneity of the catalogs, but rather to the choice
of the group-selection algorithm, the percolation methods giving
larger masses than the hierarchical one.

Here, we consider the three group catalogs of G93, each having about
450--500 groups with $cz\le 5500$ \kss. For the determination of the
MF, we retain only groups having $|b|\ge 20^{\circ}$, in order to
avoid regions of high galactic extinction.  The homogeneity of the
three catalogs, which come from the same parent galaxy sample, allows
us to cast light on the compatibility of different selection
algorithms. If we take only the nearby groups ($cz\le 2000$ \kss, i.e.
117, 124, and 136 groups) for a better comparison with P92, we find no
difference in MF among the three catalogs. This probably occurs
because G93 chose the free parameters of the selection algorithms so
as to obtain similar catalogs of groups.

However, if we consider the whole catalogs (for which statistics is
better, in particular at high masses) we find that percolation and
hierarchical algorithms give really different MFs, the former
providing larger masses. A similar difference is also reported by P92
and it is proved here to be clearly due to differences in the
algorithms. Indeed, it has been suggested that the drawback of
percolation methods is the inclusion in the catalogs of possible
non--physical systems, like a long galaxy filament aligned close to
the line--of--sight, which give large mass estimates, while the
drawback of hierarchical methods is the splitting of galaxy clusters
into various subunits, which give small mass estimates (e.g.,
Gourgoulhon et al. 1992).

The difference among the three MFs is particular relevant in
the high-mass range and leads to a flatter MF in the case of P
groups. This effect is not seen in the analysis of P92 since they have
no groups with $M> 4\times 10^{14}$\msun and less than 10 groups for
$M> 10^{14}$\msunn.

As for the completeness, the inspection of the three differential MFs
indicates that our samples are complete for $M> 9\times
10^{12}$\msunn. Similarly, P92 assumed that the selection functions of
both algorithms are efficient for $M\gtrsim 1.1\times 10^{13}$\msunn.

As for the theoretical comparisons, P92 fitted their data to
Press--Schechter predictions by assuming a unique power--law for the
fluctuation power spectrum. However, since P92 did not give the amplitude
of the fitted MF, a quantitative comparison with our results is not
straightforward. We only note here that their MF power--law slope lies
in the range $1.5\lesssim \alpha\lesssim 1.7$, being thus consistent with
our result from the Schechter--like fit, $1.3\lesssim \alpha\lesssim
1.6$. As for the value of $M_*$, P92 determined its value with
quite large uncertainties, due to the small number of high--mass
systems in their sample.
\section{SUMMARY AND CONCLUSIONS}
We analyze the three catalogs of nearby loose groups by Garcia
(1993).  This author identified groups in a magnitude--limited
redshift galaxy catalog, which covers about $\sim 2/3$ of sky within
$cz = 5500$ \kss, by using two methods, a percolation and a
hierarchical method. She tuned the free parameters of the
group-selection algorithms in order to obtain two similar catalogs of
groups and proposed a third catalog of groups defined as a combination
of the two.  Each catalog contains $\sim 450$--$500$ groups.

In agreement with previous works on earlier catalogs we find that
these groups can be described as collapsing systems. Their typical
sampled size is considerably larger than their expected virialized
region.  For all groups we compute the virial mass and correct this
mass by taking into account the young dynamical status of these
groups. We estimate corrected group masses, $M$, for two reference
cosmological models, a flat one with $\Omega_0=1$ for the matter
density parameter and an open one with $\Omega_0=0.2$.  For each of
the three catalogs we calculate the group mass function, MF.

Our main results are the following:
\begin{enumerate}
\item 
The number density of groups is not very sensitive to the choice of
the group--identification algorithm. In fact, we find that the density
of groups with $M> 9 \times 10^{12}$ \msunn, which is the adopted
limit of sample completeness, ranges in the interval $1.3$--$1.9
\times 10^{-3} (h^{-1}Mpc)^{-3}$ for $\Omega_0=1$ 
and it is about a factor of $15\%$ lower for $\Omega_0=0.2$.
\item As for the MF shape, the percolation catalog
gives a flatter MF than other catalogs.
The difference decreases 
if we do not consider the most massive groups, for which reliable
results come from galaxy cluster analysis.
\item  
After obtaining the masses contained within the central, presumably
virialized, group region by adopting a reduction in mass of $\sim
30$--$40\%$, we do a comparison with results coming from the virial
analysis of nearby rich galaxy clusters (Girardi et al. 1998a).  All
three group MFs can be regarded as a smooth extrapolation of the
cluster MF at $M<4\times 10^{14}$\msunn, which is the completeness
limit of the cluster sample.  Following a maximum--likelihood approach
and taking into account mass uncertainties, we fit the Schechter
expression for our MF on the whole mass range (groups with $9\times
10^{12}$\msun $<M< 4 \times 10^{14}$ \msun and clusters with $M>4
\times 10^{14}$ \msunn) and we obtain a slope of $\sim -1.5$ and a
characteristic mass of $\sim 3\times 10^{14}$ \msunn.
\end{enumerate}
Our result strengthens the growing evidence in favor of the continuity
of clustering properties from poor groups to very rich clusters (e.g.,
Ramella et al. 1989; Burns et al. 1996; Mulchaey \& Zabludoff 1998;
Ramella et al. 1999; Girardi et al. 2000).

Our resulting mass function of galaxy systems, which extends over two
orders of magnitude, seems to be reasonably described by
Press--Schechter predictions of models which at larger masses
describe the rich cluster mass function (in particular, in the case of
the open model).

The analysis of wide forthcoming group catalogs, e.g. the UZC (Updated
Zwicky Catalog) groups by Pisani et al. (2000) and the NOG (Nearby
Optical Galaxies) groups by Giuricin et al.  (2000), as well as
extensive spectroscopy observations of individual groups
(e.g., Zabludoff \& Mulchaey 1998; Mahdavi et al. 1999), will be of
great aid in improving the determination of the group MF, in
particular at the low--mass range which is affected by larger
uncertainties.

Furthermore, future studies of N-body cosmological simulations for
different cosmological scenarios will improve the understanding of
non-virialization effects on the estimate of group masses (e.g.,
Diaferio et al. 1999).
\acknowledgments 
We thank the referee for several suggestions.
 
We are grateful to Stefano Borgani, Marino Mezzetti, Armando Pisani
for their valuable comments, and, in particular, Massimo Ramella who
also provided us with the data of the CfA2N groups in advance of their
publication.  Special thanks to Dario Giaiotti for his help in the
initial phase of this project.  This work has been partially supported
by the Italian Ministry of University, Scientific Technological
Research (MURST), and by the Italian Space Agency (ASI).
\end{multicols}

\end{document}